%% file: main.tex
\documentclass[prd,amsmath,twocolumn,floatfix,amssymb, preprintnumbers, linenumbers,nofootinbib, superscriptaddress]{revtex4} 

\usepackage{epsf,graphicx,xcolor,amsmath}

\newcommand{\beq}{\begin{equation}}
\newcommand{\eeq}{\end{equation}}
\newcommand{\bea}{\begin{eqnarray}}
\newcommand{\eea}{\end{eqnarray}}

\begin{document}

\title{$\Upsilon$ photo-production on the proton at the Electron-Ion Collider}
\author{Oleksii Gryniuk}
\affiliation{Institut f\"ur Kernphysik \& PRISMA$^+$  Cluster of Excellence, Johannes Gutenberg Universit\"at,  D-55099 Mainz, Germany}
\author{Sylvester Joosten}
\affiliation{Argonne National Laboratory, Lemont, IL 60439, USA}
\author{Zein-Eddine Meziani}
\affiliation{Argonne National Laboratory, Lemont, IL 60439, USA}
\author{Marc Vanderhaeghen}
\affiliation{Institut f\"ur Kernphysik \& PRISMA$^+$  Cluster of Excellence, Johannes Gutenberg Universit\"at,  D-55099 Mainz, Germany}
\noaffiliation
\date{\today}

\begin{abstract}

We present a dispersive analysis with the aim to extract the $\Upsilon$-p scattering length from $\gamma p \to \Upsilon p$ experiments. In this framework, the imaginary part of the $\Upsilon$-p forward scattering amplitude is obtained from $\gamma p \to \Upsilon p$ cross section measurements, and is constrained at high energies from existing HERA and LHC data. Its real part is calculated through a once-subtracted dispersion relation, and    
the subtraction constant is proportional to the $\Upsilon$-p scattering length. We perform a feasibility study for $\Upsilon$ photo-production experiments at an Electron-Ion Collider and discuss the sensitivity and precision that can be reached in the extraction of the $\Upsilon$-p scattering length. 

\end{abstract}

\maketitle


\section{Introduction}

The interaction between heavy quarkonia, such as $J/\psi$ and $\Upsilon$, and light hadrons or nuclei provides a unique window on the gluonic van der Waals interaction in Quantum ChromoDynamics (QCD). Being a small sized system, the heavy quarkonium $Q \bar Q$ can be treated as a color dipole, and the effective two-gluon exchange interaction between the quarkonium and the light hadron  or nucleus may be estimated from the knowledge of its chromo-electric polarizability, see 
Refs.~\cite{Kharzeev:1995ij,Voloshin:2007dx,Hosaka:2016ypm} for reviews and references therein. 
Provided this effective interaction is strong enough, 
a bound state between the
$Q \bar Q$ state and the light hadron or  nucleus may be formed~\cite{Brodsky:1989jd, Wasson:1991fb, Luke:1992tm}. 
Early calculations for the chromo-electric polarizability,  treating the heavy quarkonium as a Coulombic bound state~\cite{Peskin:1979va,Bhanot:1979vb}, 
yielded estimates for the quarkonium binding energy in 
nuclear matter $B_{J/\psi} \sim 10$~MeV for $J/\psi$, and $B_{\Upsilon} \sim 2 - 4$~MeV for $\Upsilon$ ~\cite{Luke:1992tm}. Many follow-up calculations have explored the possibility of quarkonium nuclear bound states within different theoretical  frameworks~\cite{Brodsky:1997gh, Ko:2000jx, Tsushima:2011kh, Yokota:2013sfa, Beane:2014sda, Krein:2019gcm, Mamo:2019mka}. 

In recent years, the study of the excitation spectrum in the charmonium and bottomonium sectors above open charm and  open bottom thresholds has revealed a plethora of new  states, which cannot be explained as conventional $Q \bar Q$ bound states, see e.g. \cite{Olsen:2017bmm} for a recent experimental review. Several explanations for the nature of these exotic states have been put forward, ranging, among others, from tetraquark states based on QCD diquarks, QCD hybrids, hadronic molecules, or hadro-charmonium states. In contrast to conventional $Q \bar Q$ states above open charm or open bottom thresholds, for which the branching fractions in open flavor decay modes are found to be 2 or 3 orders of magnitudes larger than their hidden flavor decay modes, many of the newly found exotic states have in common that hidden flavor decay modes are discovery channels, and are only suppressed by a factor of 10 or less relative to the open flavor 
decay modes. The understanding of the nature of these states may therefore shed another light on how hidden charm or hidden bottom systems interact with light quark systems. This is especially prominent in the hadro-quarkonium models for the exotic hadrons~\cite{Dubynskiy:2008mq}, in which 
the charm or bottom $Q \bar Q$ pair remains tightly bound while interacting with the light quarks through a van der Waals interaction.

Also in the baryon sector, narrow resonances involving two heavy quarks have been discovered in recent years. In the weak decay process
$\Lambda_b \to J/\psi p K^-$, the LHCb Collaboration~\cite{Aaij:2015tga, Aaij:2019vzc} 
has found evidence for such states in the  $J/\psi p$ mass spectrum, and interpreted them as hidden-charm pentaquark states. 
As two of these states were found approximately 5 MeV and 2 MeV below the $\Sigma_c^+ \bar D^0$ and $\Sigma_c^+ \bar D^{\ast 0}$ thresholds respectively, these states were interpreted in various studies as loosely bound meson-baryon molecular states through $\pi$- or $\rho$-exchange interactions (see e.g. \cite{Roca:2015dva} among many others).  
Alternatively, the two narrow pentaquark states near $\Sigma_c^+ \bar D^{\ast 0}$ threshold, $P_c(4440)^+$ and $P_c(4457)^+$ were predicted in \cite{Eides:2015dtr} as the $1/2^-$ and $3/2^-$ hyperfine partners of deeply bound hadro-charmonium states of $\psi(2S)$ and the proton, while the narrow pentaquark state $P_c(4312)^+$ near $\Sigma_c^+ \bar D^0$ threshold was interpreted as a $1/2^+$ hadro-charmonium state of the $\chi_{c0}(1P)$ and the proton~\cite{Eides:2019tgv}. In such picture, the binding is due to the two-gluon exchange interaction between a compact quarkonium state within a proton, and is proportional to the chromo-electric polarizability of the quarkonium state. 
The hadro-charmonium framework therefore relates the nature of such exotic states involving heavy quarks to the interaction of the heavy quarkonia with light hadrons. 

The interaction of heavy quarkonia with light hadrons can also be studied within lattice QCD. 
Recently, the HAL QCD Collaboration~\cite{Sugiura:2019pye}  has performed improved lattice QCD studies of the s-wave effective potentials for 
the  $J/\psi$-nucleon system ($J = 1/2$ and $J = 3/2$), although still for an unphysical pion mass value of $m_\pi = 700$~MeV. For the $J/\psi$-nucleon system, the potential was found to be attractive, but not strong enough to allow for bound states. The lattice study extracted $J/\psi$-nucleon scattering lengths for both spin states: 
$a_{\psi p}(J = 1/2) = 0.66 \pm 0.07$~fm, and 
$a_{\psi p}(J = 3/2) = 0.38 \pm 0.05$~fm, indicating that the $J/\psi$-N state with 
$J = 1/2$  obtains significantly stronger attraction than the $J = 3/2$ state. 

In order to access the $J/\psi$-nucleon interaction from experiment, a phenomenological analysis of the $J/\psi$-p forward scattering amplitude within a dispersive framework was performed in~\cite{Gryniuk:2016mpk}. It related the imaginary part of the $J/\psi$-p forward scattering amplitude 
to $\gamma p \to J/\psi p$ and $\gamma p \to c \bar c X$ cross section data, and calculated the real part through a once-subtracted dispersion relation, for which the subtraction constant is directly related to the scattering length, and was fitted to the available data in the threshold region.   
This dispersive framework extracted as value for the spin-averaged s-wave $J/\psi$-p scattering length $a_{\psi p} = 0.046 \pm 0.005$~fm, which can be translated into a 
$J/\psi$ binding energy in nuclear matter 
of $B_{\psi} = 2.7 \pm 0.3$~MeV. Such value for the scattering length is at the lower end of the range of values estimated in the literature, ranging from 
$a_{\psi p} = 0.05$~fm~\cite{Kaidalov:1992hd} to 
$a_{\psi p} = 0.37$~fm~\cite{Sibirtsev:2005ex}. However, as the current data base for the $J/\psi$ photo-production in the threshold region is quite scarce, a reliable extraction clearly calls for new high statistics data in that region.  

A dedicated experimental program to measure the  photo-production of $J/\psi$ near threshold  has started in recent years at Jefferson Lab. The GlueX collaboration measured $J/\psi$ photo-production near threshold using the GlueX detector in Hall D and published its first  results~\cite{Ali:2019lzf}. Experiment E12-16-007~\cite{Meziani:2016lhg} in Hall C was designed and performed as a direct search of the higher mass narrow width pentaquark state $P_c^+(4450)$ in photo-production, and will provide good differential cross sections in four-momentum transfer $t$. It ran during the spring of 2019 and preliminary results are expected soon~\cite{Joosten:2020}, while experiment E12-12-001~\cite{CLAS12-tcs:proposal} in Hall B has collected data using the CLAS12 detector and a hydrogen target during  2018/2019, with the analysis presently still underway. Furthermore, a more dedicated program of high precision $J/\psi$ electro- and photo-production  measurements on the proton  using the Solenoidal Large Intensity Device (SoLID) is planned in Hall A. The Jefferson Lab experiment E12-12-006~\cite{SoLIDjpsi:proposal} will be able to measure both the electro-production differential cross sections in the four momentum transfer $t$, as well as determine the total cross section very close to the threshold region on a nucleon.  

One may also consider quarkonium production on a nucleon at the future Electron-Ion Collider (EIC) 
machine~\cite{Accardi:2012qut}. Here, it is worth noting that access to the threshold region of the $J/\psi$ production is not possible due to the lower limit of the center of mass energy of the machine. Therefore, it is best to consider a higher center of mass energy provided by the threshold production of $\Upsilon$, hence reachable by the current machine design. With sufficient integrated luminosity a precision measurement of the photo- and electro-production of $\Upsilon$ is possible, and gives a way to address the question of the existence of bottom pentaquarks. As the $\Upsilon$ production probes the gluon fields in the nucleon, such study will also shed light on the origin of the proton mass. Gluons are estimated to account for more than half of the proton mass due to the strong gluon chromo-electric and chromo-magnetic fields inside the proton~\cite{Ji:1994av}. 
At an EIC, the mass of the bottom quark as well as  the probe resolution in electro-production give two independent knobs in the investigation of the gluonic interaction between the $\Upsilon$ and the nucleon.  

To prepare for such a program of $\Upsilon$ production at an EIC, we will extend in the present work our previous dispersive study of $J/\psi$ photo-production to the case of $\Upsilon$ photo-production. We will construct the $\Upsilon$-proton forward scattering amplitude in Section 2, relating  
its imaginary part to $\gamma p \to \Upsilon p$ data. The real part will be calculated from a dispersion relation, involving one subtraction constant, which is directly related to the scattering length. 
In Section 3, we will constrain the high-energy region from existing HERA and LHC data. As no data are available so far in the threshold region, we will consider several scenarios for the scattering length providing a range of estimates for the subtraction constant. 
In Section 4, we will show results of a feasibility study for $\Upsilon$ photo-production at the EIC, considering different beam settings, and discuss the sensitivity and precision in the extraction of the 
$\Upsilon$-p scattering length from such experiments. 
In Section 5, we present our conclusions. Some technical details on the EIC simulations are presented in Appendix~\ref{apx-evgen}. 

\section{$\Upsilon$-proton forward scattering amplitude}

We consider the spin-averaged $\Upsilon p \to \Upsilon p$ forward elastic scattering process, which is described by an invariant amplitude $T_{\Upsilon p}$, depending on the crossing-symmetry variable $\nu$. The latter is defined in terms of the Mandelstam invariant $s =W^2$ as:
\bea
\nu = \frac{1}{2} (s - M^2 - M_\Upsilon^2),
\eea
where $M$ and $M_\Upsilon$ stand for the masses of the proton and $\Upsilon(1S)$ state, respectively.  
The forward differential cross section for the $\Upsilon \, p \to \Upsilon \, p$ scattering process can then be expressed as:
\bea
\frac{d \sigma}{dt} \biggr|_{t = 0} (\Upsilon p \to \Upsilon p) = \frac{1}{64 \, \pi \, s \, q_{\Upsilon p}^2} \, \big| T_{\Upsilon p}(\nu) \big|^2,
\eea
where in the forward direction the momentum transfer $t = 0$, and where $q_{\Upsilon p}$ denotes the magnitude of the $\Upsilon$ three-momentum in the c.m. frame, given by:
 \bea 
 q_{\Upsilon p}^2  = \frac{1}{4 s} \left[ s - (M_\Upsilon + M)^2 \right] \left[ s - (M_\Upsilon - M)^2 \right].
 \eea 
 
The imaginary part of the amplitude $T_{\Upsilon p}$ can be obtained as sum of elastic and inelastic discontinuities:
\bea
\Im T_{\Upsilon p}(\nu)  &=& \theta(\nu - \nu_{el}) \,  {\rm Disc}_{\rm el} T_{\Upsilon p}(\nu) \nonumber \\
&+&   \theta(\nu - \nu_{inel}) \,  {\rm Disc}_{\rm inel} T_{\Upsilon p}(\nu).
\label{eq:disctot}
\eea
The elastic discontinuity starts from elastic threshold $s = s_{el} = (M_\Upsilon + M)^2 = 108.13$~GeV$^2$, or equivalently $\nu_{el} = M_\Upsilon M = 8.88$~GeV$^2$, whereas the inelastic discontinuity starts at the $B \bar B$ meson production threshold, corresponding with $s_{inel} = (M + 2 M_B)^2 = 132.18$~GeV$^2$, or equivalently $\nu_{inel} = 20.90$~GeV$^2$. 

Analogous to the $J/\psi$ case~\cite{Gryniuk:2016mpk}, we will parametrize the elastic and inelastic discontinuities of the $\Upsilon p$  forward scattering amplitude by the following 3-parameter forms, for $x = {\rm el/inel}$:
\bea
{\rm Disc}_{x} T_{\Upsilon p}(\nu)  &=& 
C_{x} \left( 1 - \frac{\nu_{x}}{\nu} \right)^{b_{x}}  \left( \frac{\nu}{\nu_{x}} \right)^{a_{x}} ,
\label{eq:discx} 
\eea
where the factors $\sim (1 - \nu_x / \nu)^{b_x}$  
determine the behavior around the respective threshold $\nu_x$, and the 
factors  $\sim \nu^{a_x}$ determine the Regge behavior of the amplitude at large $\nu$. 
In the following we will discuss how we can determine the respective parameters 
appearing in the elastic and inelastic discontinuities. 

For the discontinuity across the elastic cut, 
${\rm Disc}_{el} T_{\Upsilon p}$, we  use the vector meson dominance (VMD) assumption to relate the $\Upsilon p$ elastic cross section 
$\sigma_{\Upsilon p}^{el}$ to the total $\gamma p \to \Upsilon p$ photo-production cross section~\cite{Barger:1975ng,Redlich:2000cb}: 
\bea
{\rm Disc}_{\rm el} T_{\Upsilon p}(\nu)  &=& 2 \sqrt{s} \, q_{\Upsilon p} \sigma_{\Upsilon p}^{el} 
\label{eq:sigmael}
\\
&\simeq&  2 \sqrt{s} \, q_{\Upsilon p}  \left( \frac{M_\Upsilon}{e f_\Upsilon} \right)^2 \left( \frac{q_{\gamma p}}{q_{\Upsilon p}} \right)^2 
\sigma (\gamma p \to \Upsilon p), 
\nonumber
\eea
with electric charge $e$ given through $\alpha = e^2 / (4 \pi) \simeq 1/137$, and  where $f_\Upsilon$ is the $\Upsilon$ decay constant, which is obtained from the $\Upsilon \to e^+ e^-$ decay as 
\begin{eqnarray}
\Gamma_{\Upsilon \to ee} = \frac{4 \pi \alpha^2}{3} \frac{f_\Upsilon^2}{M_\Upsilon}.
\end{eqnarray}
The experimental value $\Gamma_{\Upsilon \to ee} =  1.34 \pm 0.02$~keV~\cite{Tanabashi:2018oca} yields $f_\Upsilon = 0.238$~GeV. Furthermore, $q_{\gamma p}$ denotes the magnitude of the 
photon three-momentum in the c.m. frame of the $\gamma p \to \Upsilon p$ process:
\bea
q_{\gamma p} = \frac{(s - M^2)}{2 \sqrt{s}}.
\eea

The discontinuity across the inelastic cut, ${\rm Disc}_{inel} T_{\Upsilon p}$, 
is related through the optical theorem to the $\Upsilon  p \to b \bar b X$ inelastic cross section 
$\sigma_{\Upsilon p}^{inel}$ as:
\bea
{\rm Disc}_{\rm inel} T_{\Upsilon p}(\nu) = 2 \sqrt{s} \, q_{\Upsilon p} \, \sigma_{\Upsilon p}^{inel}.    
\label{eq:sigmainel}
\eea
Using VMD, one can relate the total $\gamma p \to b \bar b X$ photo-production cross section to the inelastic cross sections for the sum over $\Upsilon$ states (labeled by index $i$):
\begin{eqnarray}
\sigma (\gamma p \to b \bar b X) = \sum_{\Upsilon_i}
 \left( \frac{q_{\Upsilon_i p}}{q_{\gamma p}} \right)^2
 \,  \left( \frac{e f_{\Upsilon_i}}{M_{\Upsilon_i}} \right)^2  \sigma_{\Upsilon_i p}^{inel} \,  . 
\label{eq:sigmainel2}
\end{eqnarray}
In contrast to the elastic process, where the final state is fixed to be the $\Upsilon(1S)$-p state, and for which the elastic photo-production cross section $\sigma (\gamma p \to \Upsilon p)$ can be expected to be approximately dominated by its lowest term in the sum over vector bottomonia states, the open bottom $b \bar b$ final state in the inelastic photo-production cross section of Eq.~(\ref{eq:sigmainel2})  can be expected to get sizeable contributions from several vector bottomonia states. We will therefore refrain from approximating Eq.~(\ref{eq:sigmainel2}) by its lowest vector bottomonium contribution in the following analysis, and instead constrain the inelastic discontinuity through the forward differential cross section $d \sigma / dt$ for  $\gamma p \to \Upsilon p$ as discussed in the next section. 

The real part of the scattering amplitude $T_{\Upsilon p}$ is related to its imaginary part
through a once-subtracted forward dispersion relation:
\beq
\Re T_{\Upsilon p}(\nu) = T_{\Upsilon p}(0) + \frac{2}{\pi} \nu^2 \int_{\nu_{el}}^\infty d
\nu^\prime \frac{1}{\nu^\prime} \frac{\Im T_{\Upsilon p}(\nu^\prime)}{\nu^{\prime \, 2} - \nu^2},
\label{eq:disp}
\eeq
with $T_{\Upsilon p}(0) $ the subtraction constant at $\nu = 0$. In this work, the subtraction constant is suggested to be obtained by performing a fit of the 
differential $\gamma p \to \Upsilon p$  photo-production cross section data at $t=0$, which is related to $T_{\Upsilon p}$ as:
\beq
\frac{d \sigma}{dt} \biggr|_{t = 0} (\gamma p \to \Upsilon p) 
= \left( \frac{e f_\Upsilon}{M_\Upsilon} \right)^2  \frac{1}{64 \, \pi \, s \, q_{\gamma p}^2} \, \big| T_{\Upsilon p}(\nu) \big|^2.
\label{eq:dsigmadt0_gapjpsip}
\eeq

The real part of the forward scattering amplitude at threshold $T_{\Upsilon p}(\nu_{el}) $ is directly related to the $\Upsilon$-p scattering length 
$a_{\Upsilon p}$ as:
\bea
T_{\Upsilon p}(\nu = \nu_{el}) = 8 \pi (M + M_\Upsilon) \, a_{\Upsilon p}. 
\eea
Analogously to the $J/\psi$ case, 
we may relate a positive $\Upsilon$-p scattering length, corresponding to an attractive interaction, 
to an $\Upsilon$ binding energy $B_\Upsilon$ in nuclear matter, using a linear density approximation~\cite{Kaidalov:1992hd}:
\begin{eqnarray}
B_\Upsilon \simeq \frac{8 \pi (M + M_\Upsilon) a_{\Upsilon p}}{4 M M_\Upsilon} \, \rho_{nm},
\label{eq:nmbe}
\end{eqnarray}
where $\rho_{nm} \simeq 0.17$~fm$^{-3}$ denotes the nuclear matter density.

As the aim of our work is to extract $a_{\Upsilon p}$ from fitting the subtraction constant $T_{\Upsilon p}(0)$ to future $\Upsilon$ photo-production data in the threshold region, we will consider three scenarios for the subtraction constant in order to explore the data sensitivity to its extraction. 

The simplest scenario corresponds with having zero value of the subtraction constant, i.e. $T_{\Upsilon p}(0) = 0$. The real part of the $\Upsilon$-p  scattering amplitude is then fully determined by its imaginary part through the dispersion integral 
in Eq.~(\ref{eq:disp}). The resulting value of the $\Upsilon$-p scattering length is then extremely small, around $a_{\Upsilon p} \sim 10^{-3}$~fm. 

A second scenario is to estimate the subtraction constant by a scaling from its value for the $J/\psi p$ scattering, which was obtained from a fit to data in~\cite{Gryniuk:2016mpk} as $T_{J/\psi p }(0) \simeq 22.5 \pm 2.5$. 
Observing that at high energies the normalizations of both the $J/\psi p$ and $\Upsilon p$ scattering amplitudes are completely driven by their inelastic discontinuities, as discussed in the following section, and making the strong assumption that the subtraction constants scale in the same way, yields the estimate: 
\beq
T_{\Upsilon p}(0) \approx T_{J/\psi p }(0) \cdot C_{\rm inel}^\Upsilon / C_{\rm inel}^{J/\psi} \approx T_{J/\psi p }(0).
\eeq
With cross section normalization estimate 
at high energies  
$C_{\rm inel}^\Upsilon / C_{\rm inel}^{J/\psi} \approx 0.9$, which we discuss in the next section, this second scenario yields $T_{\Upsilon p}(0) = 20.5$, which corresponds with a scattering length $a_{\Upsilon p} \simeq 0.016$~fm. 

We also consider a third, theoretically more motivated, scenario, in which an estimate of the $\Upsilon$-p threshold amplitude is obtained by considering, in the leading approximation,  the heavy bottomonium 
as a Coulombic bound state which interacts with the proton through its chromo-electric polarizability~\cite{Peskin:1979va} yielding:
\beq
T_{\Upsilon p}(\nu_{el}) = \frac{16\pi^2}{9} \alpha_\Upsilon M_{\Upsilon} M^2,
\eeq
with $\alpha_\Upsilon$ the chromo-electric polarizability.  
For a Coulombic bound state, $\alpha_\Upsilon$ is given by~\cite{Peskin:1979va,Bhanot:1979vb}
\beq
\alpha_\Upsilon = \frac{28}{81} \pi a_0^3,
\label{eq:elpol}
\eeq
where $a_0$ is the Bohr radius of the $\Upsilon$ state, given by
\beq
a_0^{-1} = \frac{2}{3}\alpha_s m_b.
\eeq
Using the parameter values for the strong coupling $\alpha_s\approx 0.37$ and the bottom quark mass $m_b\approx 4.76$ GeV from a recent potential model calculation for the bottomonium spectrum~\cite{Deng:2016ktl}, the Bohr radius for the $\Upsilon$ takes on the value
$a_0^{-1} \simeq (0.85)^{-1} \,\mathrm{GeV}$. 
Eq.~(\ref{eq:elpol}) then yields a chromo-electric polarizability $\alpha_\Upsilon \simeq 0.67$~GeV$^{-3}$, in good agreement with the recent evaluation given in Ref.~\cite{Brambilla:2015rqa}: 
$\alpha_\Upsilon = 0.5 \substack{+0.42 \\ -0.38}$~GeV$^{-3}$, using color octet intermediate states in the calculation of the polarizability instead of free ones as in ~\cite{Peskin:1979va,Bhanot:1979vb}. For the purpose of providing cross section estimates in our third scenario, we will use an average between both results: $\alpha_\Upsilon \simeq 0.6$~GeV$^{-3}$, 
which yields $T_{\Upsilon p}(\nu_{el}) \simeq 88$ and for the s-wave scattering length $a_{\Upsilon p} \simeq 0.066$~fm. Using the dispersion integral of Eq.~(\ref{eq:disp}) in making the small extrapolation between the amplitudes at $\nu = \nu_{\rm el}$ and $\nu = 0$, this third scenario then yields as value for the subtraction constant: $T_{\Upsilon p}(0) \simeq 87$.

\section{Fit to existing data for $\Upsilon$ photo-production on the proton}

In this section, we discuss the fit of the elastic and inelastic discontinuities, which are parameterized according to the three-parameter forms of Eq.~(\ref{eq:discx}), to available  $\Upsilon$-p photo-production data, following similar strategy as our previous analysis for 
the $J/\psi$-p system~\cite{Gryniuk:2016mpk}. 
 
\begin{figure}[h]
\includegraphics[width=0.49\textwidth]{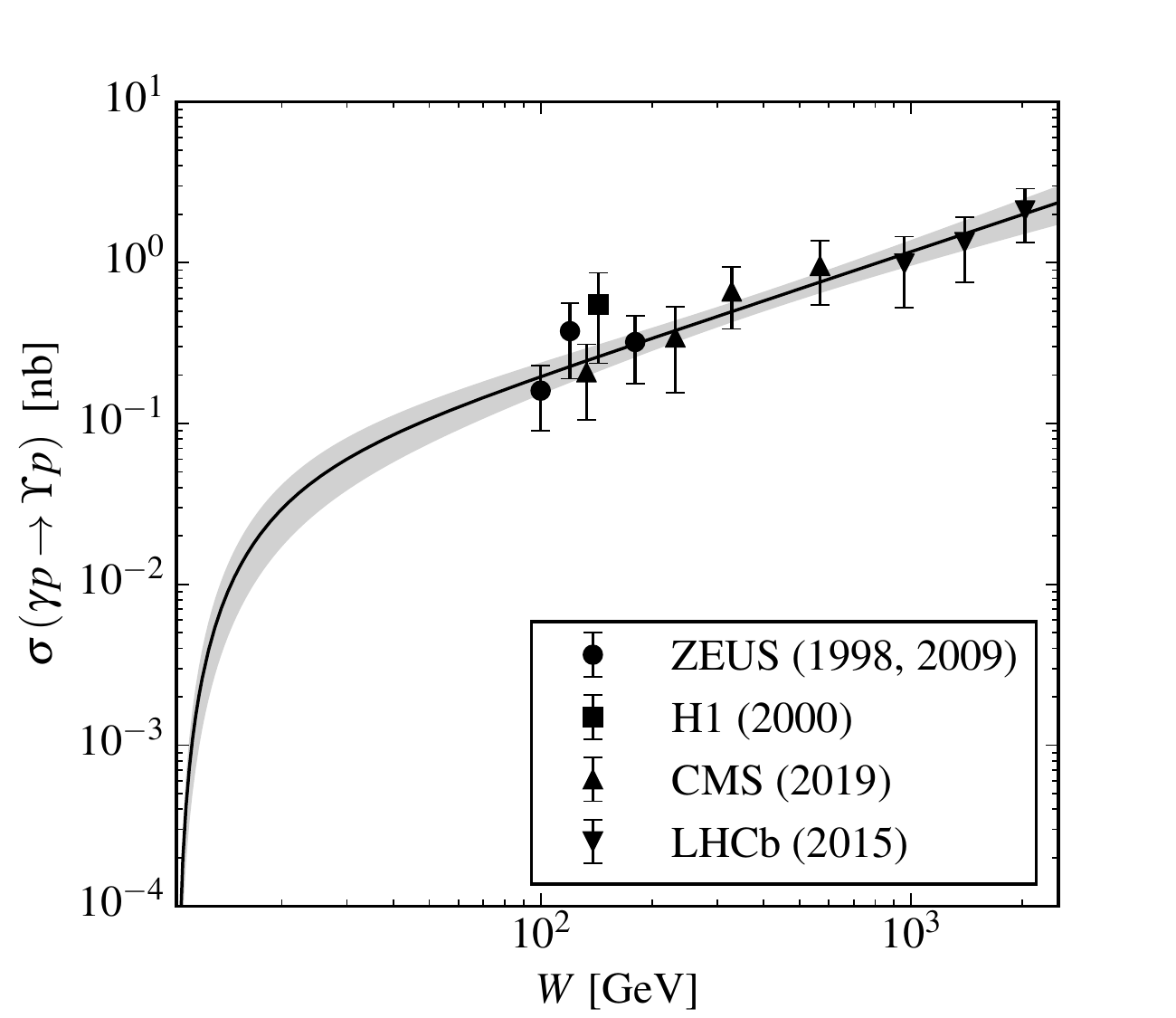}
\includegraphics[width=0.49\textwidth]{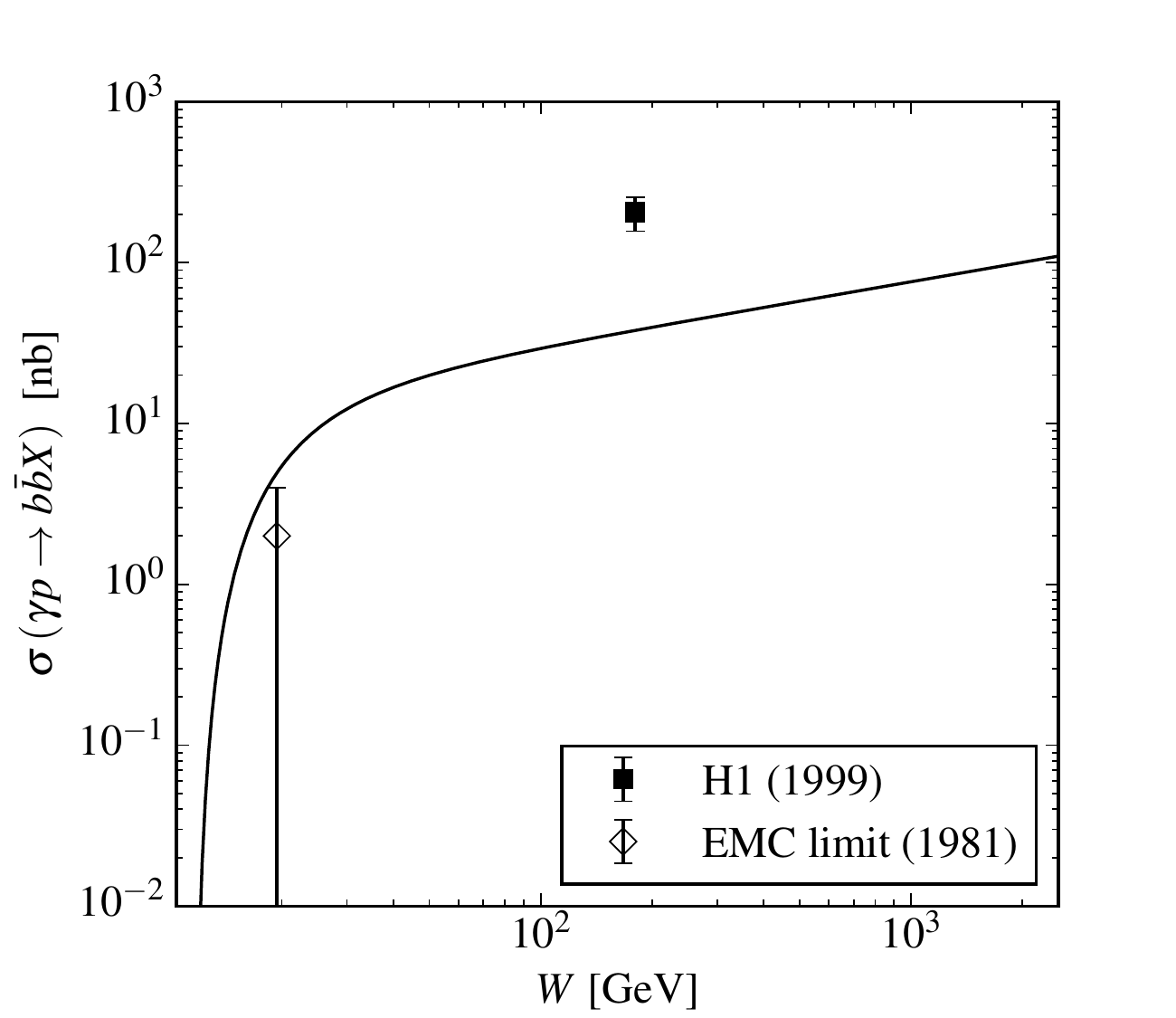}
\caption{Upper panel: $W$-dependence of the $\gamma p \to \Upsilon p$ total cross section.
The curve is the result of our fit with parameters given in Table~\ref{tab:fits}.
The elastic $\Upsilon$ photo-production data are from HERA: H1~\cite{Adloff:2000vm}
and ZEUS~\cite{Breitweg:1998ki, Chekanov:2009zz}, and from LHC: LHCb~\cite{Aaij:2015kea} 
and CMS~\cite{Sirunyan:2018sav}. 
Lower panel: $W$-dependence of the 
total inelastic photo-production cross section $\gamma p \to b \bar b X$.
The curve gives a lower limit arising from the contribution of the $\Upsilon(1S)$ state only in the sum of Eq.~(\ref{eq:sigmainel2}) for the $\gamma p \to b \bar b X$ cross section. 
The open beauty photo-production data points are from HERA~\cite{Adloff:1999nr} and EMC~\cite{Aubert:1981gx}.}
\label{fig:sigmatot}
\end{figure}

At present, the $\gamma p \to \Upsilon p$ photo-production database consists of four data points from HERA~\cite{Adloff:2000vm,Breitweg:1998ki,Chekanov:2009zz}, shown in Fig.~\ref{fig:sigmatot} 
(upper panel). 
Furthermore at Large Hadron Collider (LHC) energies, the $\gamma p \to \Upsilon p$ cross section has been extracted from central pp production data at LHCb~\cite{Aaij:2015kea} 
and from ultra-peripheral pPb collisions at  CMS~\cite{Sirunyan:2018sav}.

The inclusive $b \bar b$ photo-production database is represented by one data point from HERA~\cite{Adloff:1999nr}.
Additionally, the lower energy cross section upper limit from EMC~\cite{Aubert:1981gx}
is added to guide the low-energy behavior, as shown in Fig.~\ref{fig:sigmatot} (lower panel).

For high energies $\sqrt{s} = W\sim100$ GeV, the differential cross section has been measured by the ZEUS Collaboration and follows an exponential t-dependence:
\beq
\frac{d \sigma}{dt} (\gamma p \to \Upsilon p)
\;=\; A \cdot e^{Bt}, \quad \quad 
A = \frac{d \sigma}{dt} \biggr|_{t = 0} (\gamma p \to \Upsilon p) ,
\label{eq:bdef}
\eeq
with an empirical slope parameter $B(W = 100$~GeV$)=4.5\pm0.5$~GeV$^{-2}$~\cite{Chekanov:2009zz}. 
The exponential dependence of Eq.~(\ref{eq:bdef}) allows us to express the extrapolated value
 of the differential cross section at $t=0$ as
\beq
A  \simeq B e^{- B t_{\rm min}} \cdot\sigma (\gamma p \to \Upsilon p),
\label{eq:brel}
\eeq
where
\beq
t_\mathrm{min} = M_\Upsilon^2 - 2q_{\gamma p} \left(\sqrt{q_{\Upsilon p}^2 + M_\Upsilon^2} - q_{\Upsilon p}\right)
\eeq
is the minimum (modulo) physical momentum transfer, corresponding to the forward scattering ($\theta_{\gamma \Upsilon}=0$).

On physical grounds, one may expect the exponential dependence of Eq.~(\ref{eq:bdef}) to hold in a limited $t$-range only, turning into a power dependence at larger $t$ values. At high $W$, this only gives a minor correction to Eq.~(\ref{eq:brel}), but at lower $W$-values one may expect the correction to be more important. 
In this case, one should apply the fit form of Eq.~(\ref{eq:brel}) only in the limited $t$-range in order to extrapolate to $A$. 

The present database for $\Upsilon$ photo-production is unfortunately insufficient to perform a good fit using the forms of Eq.~(\ref{eq:discx}) with all parameters unconstrained.
Especially the lack of low-energy data prevents a direct determination of the low-energy slope parameters $b_x$ of the cross sections at present.
Assuming a similarity in the energy dependence of the 
cross sections for charm and bottom photo-production, 
we thus start by simply fixing $b_{\rm el}$ and 
$b_{\rm inel}$ to the values obtained in the $J/\psi$ analysis~\cite{Gryniuk:2016mpk}. 
The high-energy elastic slope parameter $a_{\rm el}$ and  
the elastic normalization constant $C_{\rm el}$ 
are then fitted to the available data points for the elastic $\Upsilon$ photo-production total cross section, as shown in Fig.~\ref{fig:sigmatot} (upper panel), yielding the parameter values shown in Table~\ref{tab:fits} (second column).  

\begin{table}[h]
\begin{tabular*}{\columnwidth}{c @{\extracolsep{\fill}} cccc}
\hline
\hline
& \quad $x$ = el \quad & \quad $x$ = inel \quad\\
\hline
\input{table.tex}
\hline
\hline
\end{tabular*}
\caption{Fit results for the coefficients entering the elastic discontinuity (second column, $x = {\rm el}$), 
and the inelastic discontinuity (third column, $x = {\rm inel}$).
The parameters $b_{\rm el}, b_{\rm inel}$, and 
$a_{\rm inel}$ are fixed from the $J/\psi$ analysis of Ref.~\cite{Gryniuk:2016mpk}.
}
\label{tab:fits}
\end{table}

To determine the inelastic normalization constant $C_{\rm inel}$, it was observed in \cite{Gryniuk:2016mpk} for the $J/\psi$ case that around 
$W = 100$~GeV the inelastic discontinuity is around two orders of magnitude larger than its elastic counterpart. Although there is only one data point for the inclusive $\gamma p \to b \bar b X$ cross section, shown in the lower panel of  Fig.~\ref{fig:sigmatot}, it also confirms this finding for the $\Upsilon$-p discontinuities.   
The amplitude $T_{\Upsilon p}$ entering the forward differential cross section $A$ in Eq.~(\ref{eq:brel}) is thus dominated by the normalization constant $C_{\rm inel}$. At high energies around $W \sim 100$~GeV, where the subtraction constant makes a negligible contribution to $A$, we thus solve Eq.~(\ref{eq:brel}) and fix the normalization $C_{\rm inel}$ by the available high-energy data for the $t$-slope parameter $B$.

\begin{figure}[h]
\includegraphics[width=0.49\textwidth]{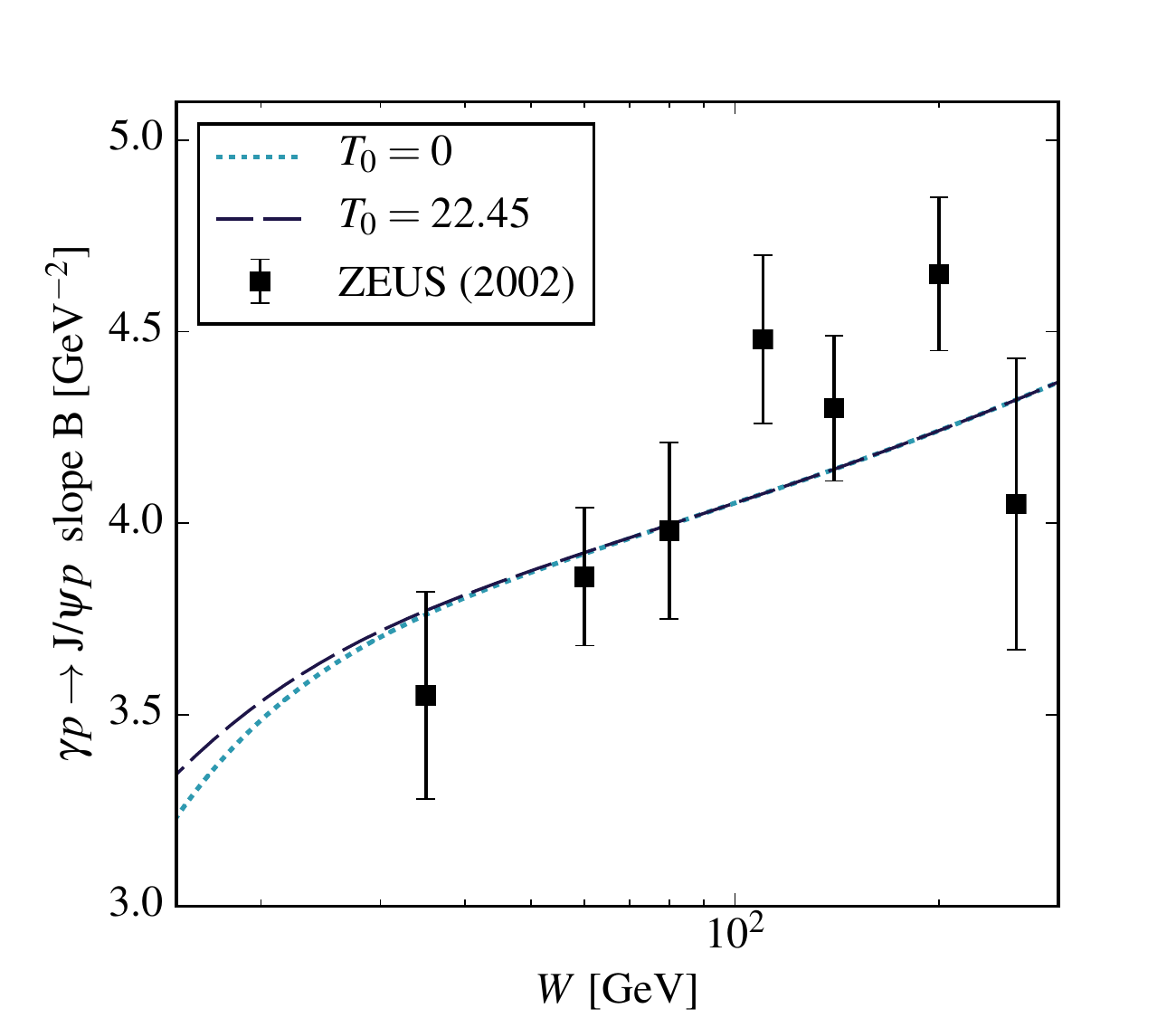}
\includegraphics[width=0.49\textwidth]{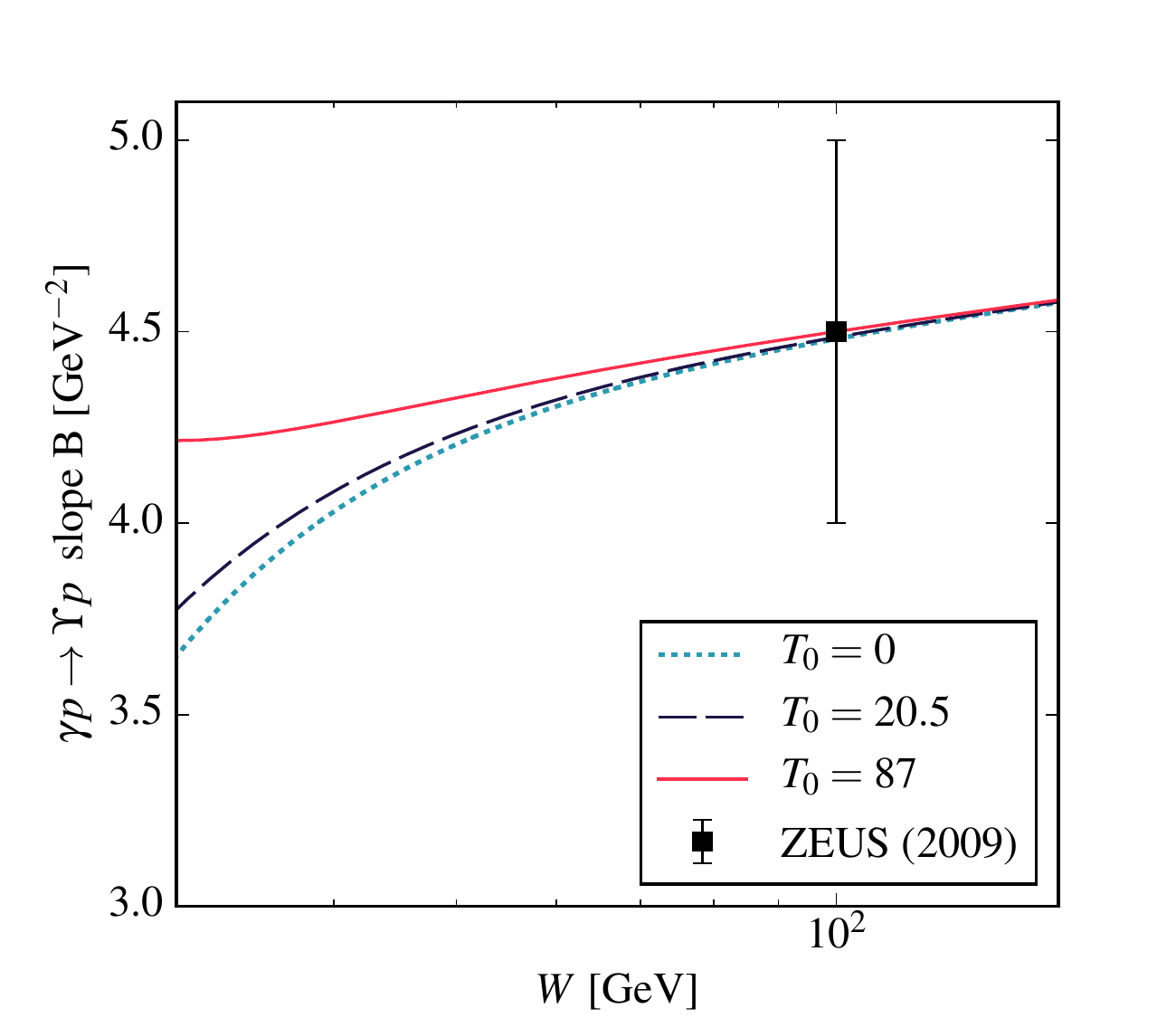}
\caption{
The $W$-dependence of the $t$-slope parameter $B$ in  Eq.~(\ref{eq:bdef}). The curves are obtained by solving 
Eq.~(\ref{eq:brel}) for different values of the subtraction constant $T_0$ in the $J/\psi p $ (upper)  and $\Upsilon p$ (lower) forward scattering amplitudes.
The $t$-slope data points are from HERA: for $J/\psi$ from~\cite{Chekanov:2002xi}, 
and for $\Upsilon$ from~\cite{Chekanov:2009zz}. 
The red solid curve, corresponding to $T_{\Upsilon p}(0)=87$, passes through the central value of the data point exactly, according to our constraint on the $C_\mathrm{inel}$.
}
\label{fig:bslope}
\end{figure}

We show this constraint based on the $t$-slope parameter $B$ in Fig.~\ref{fig:bslope}, and illustrate the calculation procedure first for the $J/\psi$ case (upper panel).
In solving Eq.~(\ref{eq:brel}) for $B$, we use the total cross section data for $\sigma(\gamma p \to J/\psi p)$ which fixes the elastic discontinuity. Furthermore, on the left hand side of Eq.~(\ref{eq:brel}), we need the real and imaginary parts of the forward scattering amplitude to determine $A$, according to Eq.~(\ref{eq:dsigmadt0_gapjpsip}). 
For the imaginary part of the amplitude, which is described by its elastic and inelastic discontinuities, we use $C_{\rm inel}$ and $a_{\rm inel}$ as fit parameters.   
The real part of the amplitude is calculated from the dispersion relation of Eq.~(\ref{eq:disp}) for three values of the subtraction constant $T_0 \equiv T_{\psi p}(0)$, considered in Ref.~\cite{Gryniuk:2016mpk}. The solution of Eq.~(\ref{eq:brel}) for these three values of $T_0$ is shown in Fig.~\ref{fig:bslope} and compared with the HERA data for the $t$-slope. One firstly sees from Fig.~\ref{fig:bslope} that for $W \geq 50$~GeV, the sensitivity to the subtraction constant becomes vanishingly small, thus allowing to determine $a_{\rm inel}$ and $C_{\rm inel}$ from a fit to the HERA data as: $a_{\rm inel} = 1.20$ and 
$C_{\rm inel} = 20.5$.

We apply the same procedure to the $\Upsilon$ $t$-slope parameter in the lower panel of Fig.~\ref{fig:bslope}, and  show our results for the three scenarios for the subtraction constant $T_0 \equiv T_{\Upsilon p}(0)$ discussed above. We again observe that for $W \geq 100$~GeV, the sensitivity to the subtraction constant becomes vanishingly small. 
As there is only one data point at $W = 100$~GeV in this case, we fix $a_{\rm inel}$ to the $J/\psi$ value and extract $C_{\rm inel}$ by constraining the $t$-slope to the data point at 100 GeV. The obtained value for $C_{\rm inel}$ is listed in Table~\ref{tab:fits}. It is seen that the value of the thus extracted dimensionless parameter $C_{\rm inel}$ is similar within 10 \% between the $J/\psi$ and $\Upsilon$ cases. We also note that in the sum of Eq.~(\ref{eq:sigmainel2}) the $\Upsilon$ contribution  to the inelastic open beauty photo-production cross section is around 20~\%. 

\begin{figure}
\includegraphics[width=0.49\textwidth]{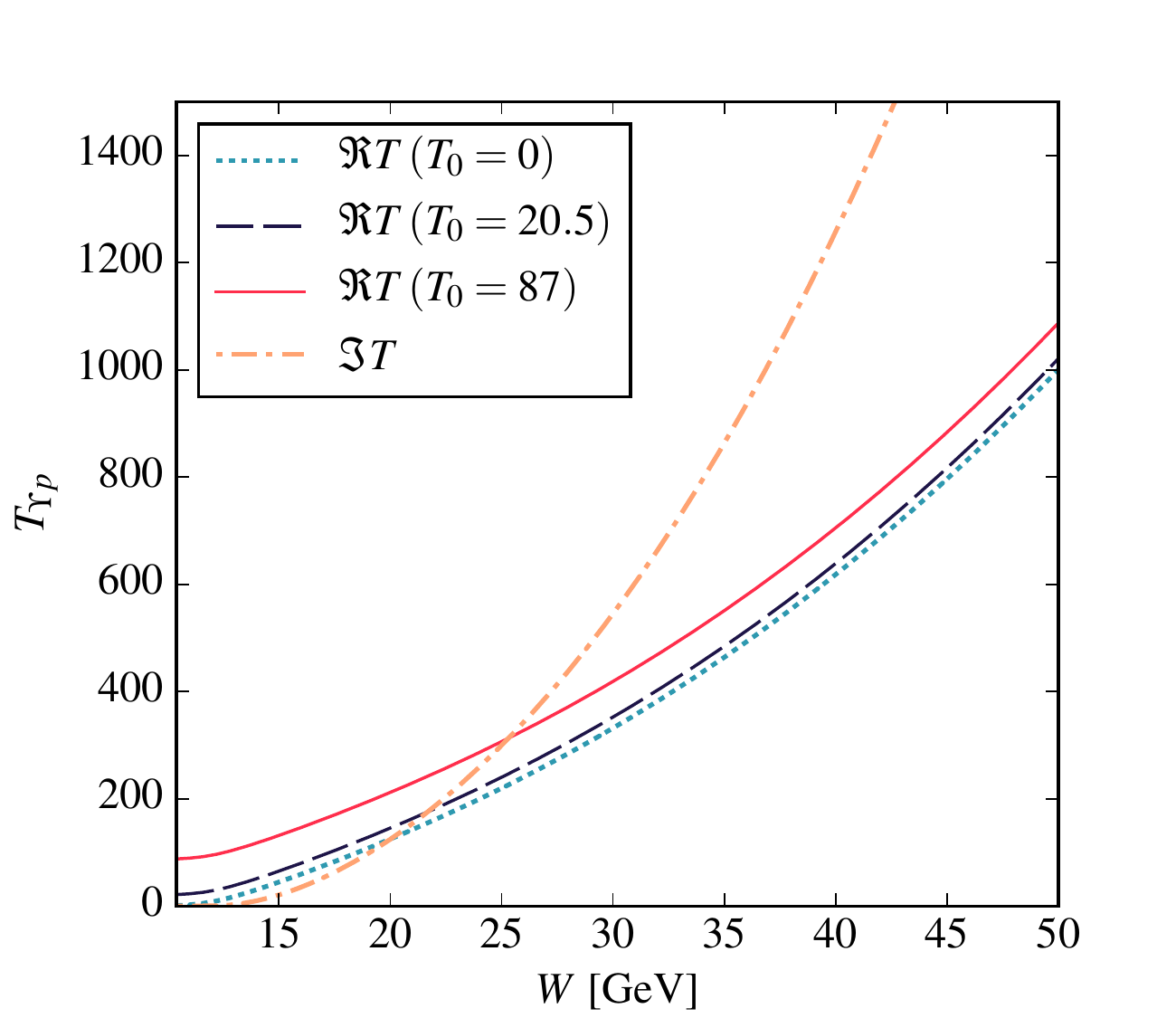}
\caption{
Imaginary part (dash-dotted curve) and real part of the forward scattering amplitude $T_{\Upsilon p}$ as function of $W$.
The real part is shown for different values of the subtraction constant as indicated on the figure.
}
\label{fig:psip_psip}
\end{figure}

In Fig.~\ref{fig:psip_psip}, we show the $W$-dependence of real and imaginary parts of the scattering amplitude $T_{\Upsilon p}$ in our dispersive formalism, for the three choices of the subtraction constant discussed above. We notice that the real part of the amplitude dominates in the threshold region, whereas the imaginary part dominates at high energies as expected for a diffractive process. For the largest value of the subtraction constant $T_0 = 87$ considered, the imaginary part overtakes the real part around $W \approx 25$~GeV.

\section{Results for $\Upsilon$ photo-production at the EIC and discussion}

We investigate in this section how to extract the subtraction constant from a fit to the 
 differential $\gamma p \to \Upsilon p$ cross section data at an Electron-Ion Collider (EIC). 
 
We consider both a medium-energy and high-energy EIC configuration. 
The medium-energy configuration (setting 1) has a 10~GeV electron beam incident on a 100~GeV proton beam 
($\sqrt{s_{ep}} = 63$~GeV), while the high-energy configuration (setting 2) has a 18~GeV electron beam incident 
on a 275~GeV proton beam ($\sqrt{s_{ep}} = 140$~GeV), 
corresponding to nominal configurations for the EIC design.

For the $\gamma p \to \Upsilon p$ cross section, we use the dispersive model discussed above with $t$-dependence as in Eq.~(\ref{eq:bdef}). We show results for the three scenarios for the subtraction constant $T_{\Upsilon p}(0)$, and corresponding s-wave scattering length $a_{\Upsilon p}$ in Table~\ref{tab:scattlength}. 
The uncertainties correspond to the simulated EIC 
$\gamma p \to \Upsilon p$ data for the two beam settings.

\begin{table}[h]
\begin{tabular*}{\columnwidth}{c @{\extracolsep{\fill}} ccc}
\hline
\hline
\quad setting \quad & \quad $T_{\Upsilon p}(0)$ \quad &  \quad $a_{\Upsilon p}$ (in fm) \quad  & \quad $B_{\Upsilon}$ (in MeV) \quad \\ 
\hline
1 \input{table2_eic1.tex}
\hline
2 \input{table2_eic2.tex}
\hline
\hline
\end{tabular*}
\caption{Values of 
the subtraction constant 
$T_{\Upsilon p}(0)$ (second column), 
the corresponding $\Upsilon$-p s-wave scattering length $a_{\Upsilon p}$ (third column), 
and the corresponding $\Upsilon$-nuclear matter binding energy $B_\Upsilon$, according to Eq.~(\ref{eq:nmbe}) (fourth column).
The uncertainty estimates are propagated based on the generated EIC differential cross section data points.
}
\label{tab:scattlength}
\end{table}

\begin{figure*}
\includegraphics[width=0.49\textwidth]{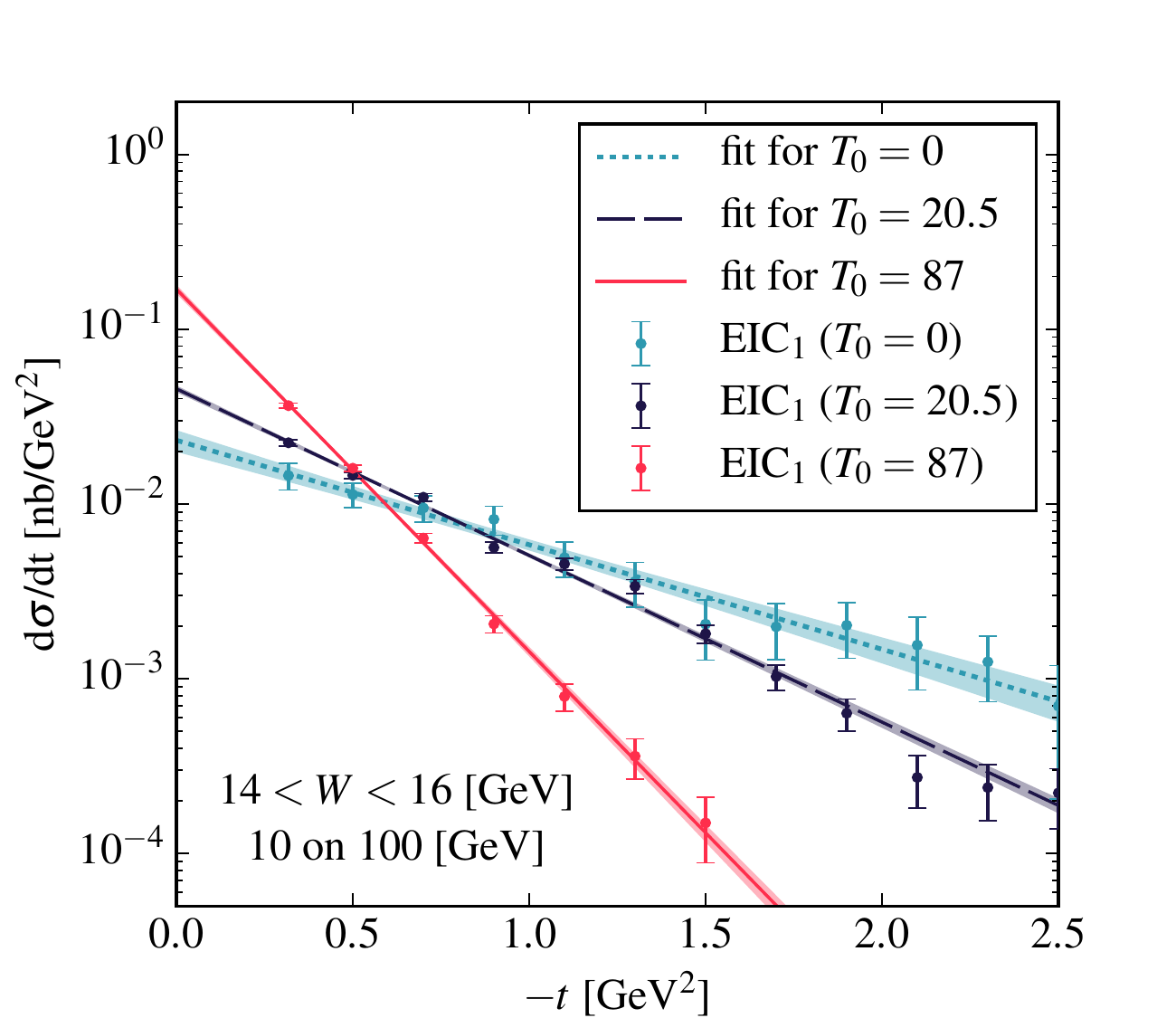}
\includegraphics[width=0.49\textwidth]{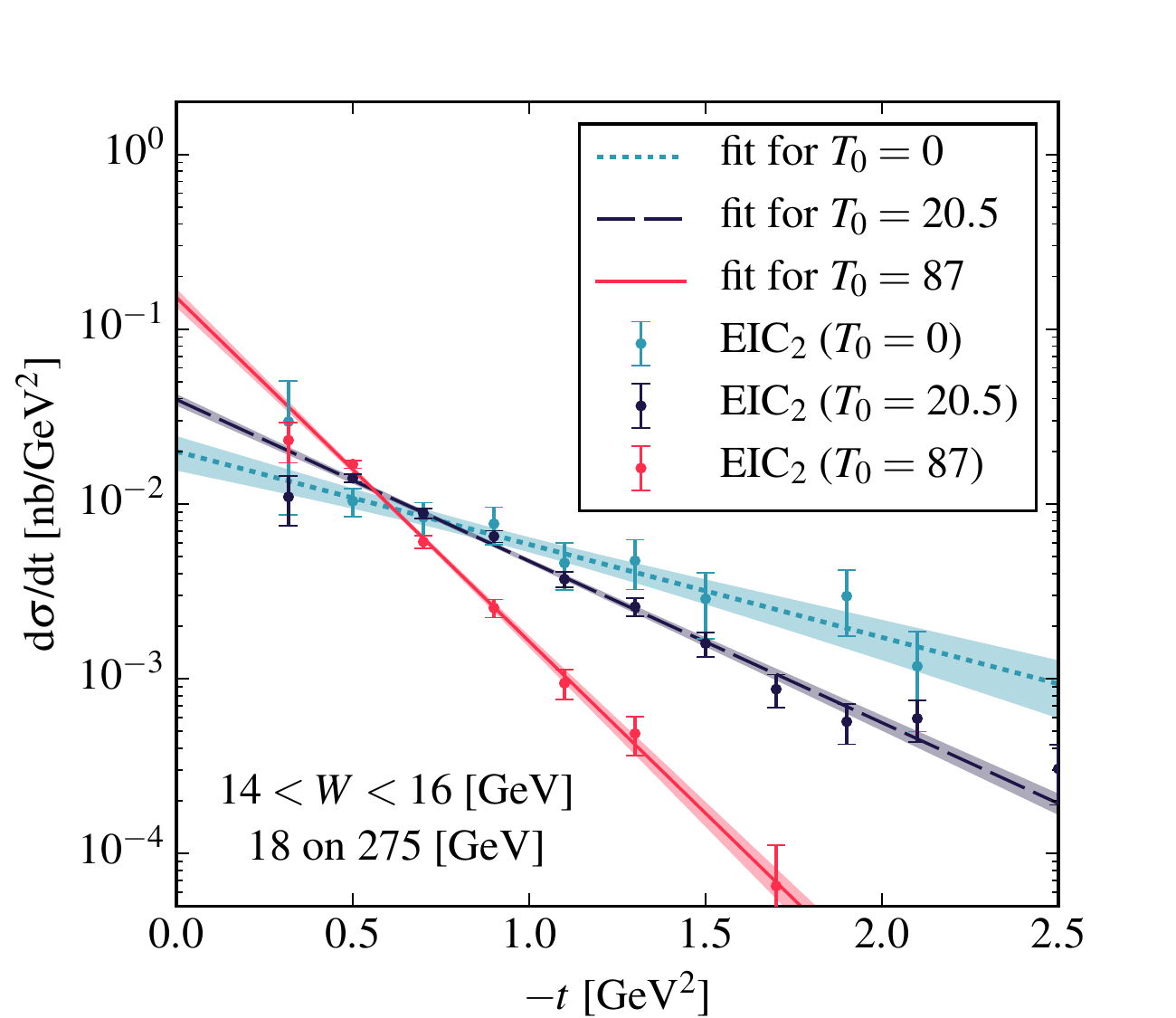} \\
\includegraphics[width=0.49\textwidth]{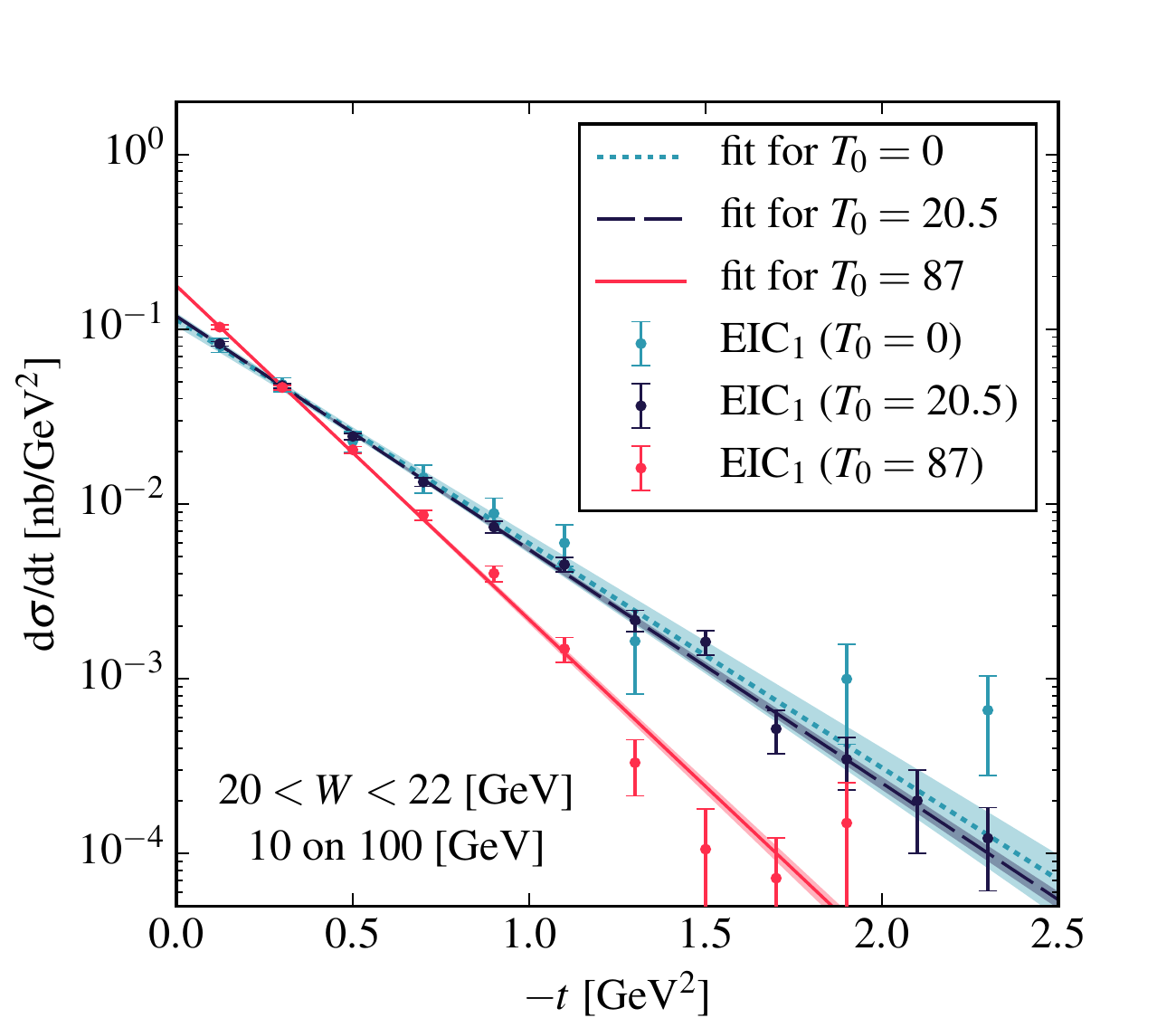}
\includegraphics[width=0.49\textwidth]{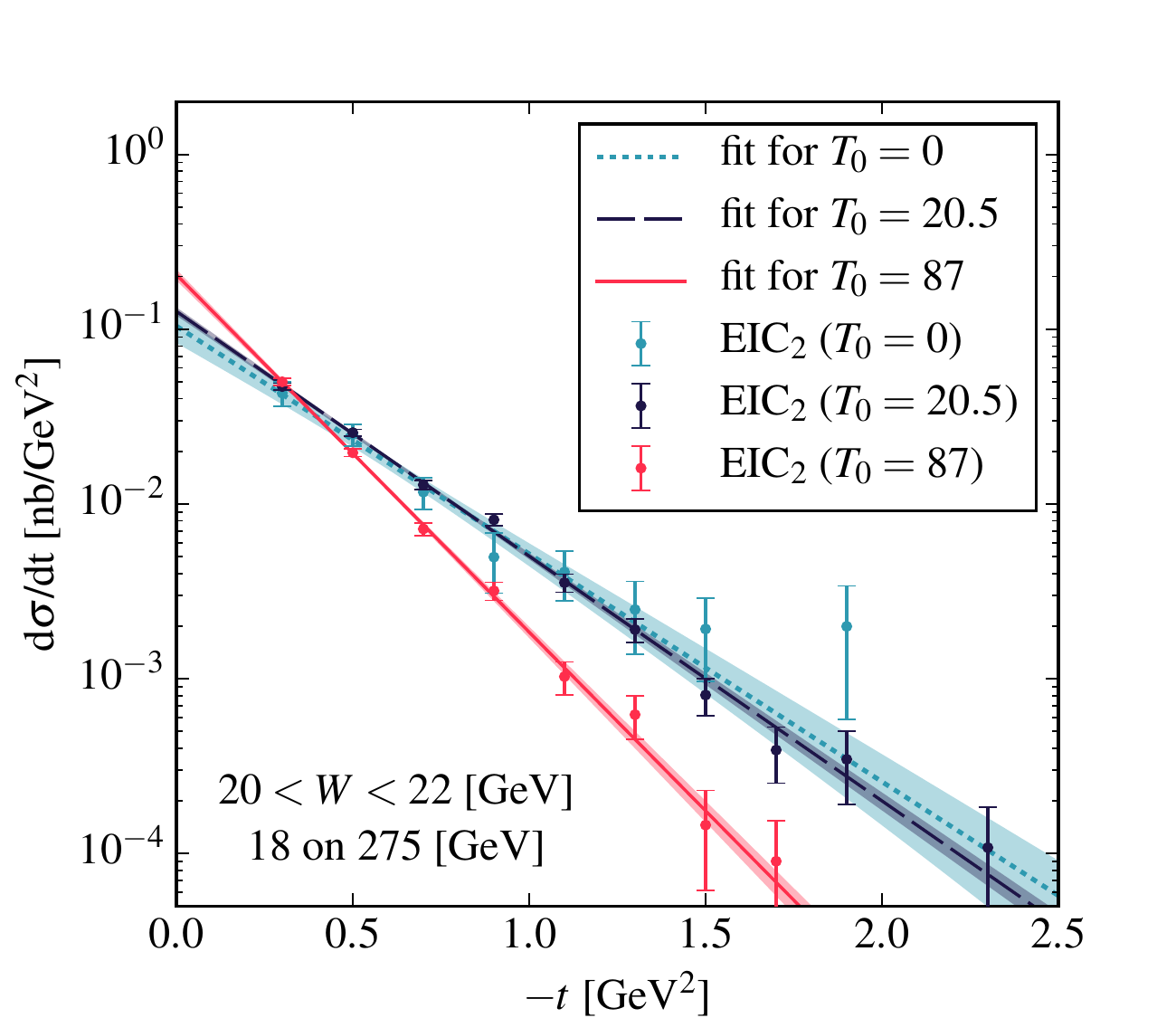} \\
\includegraphics[width=0.49\textwidth]{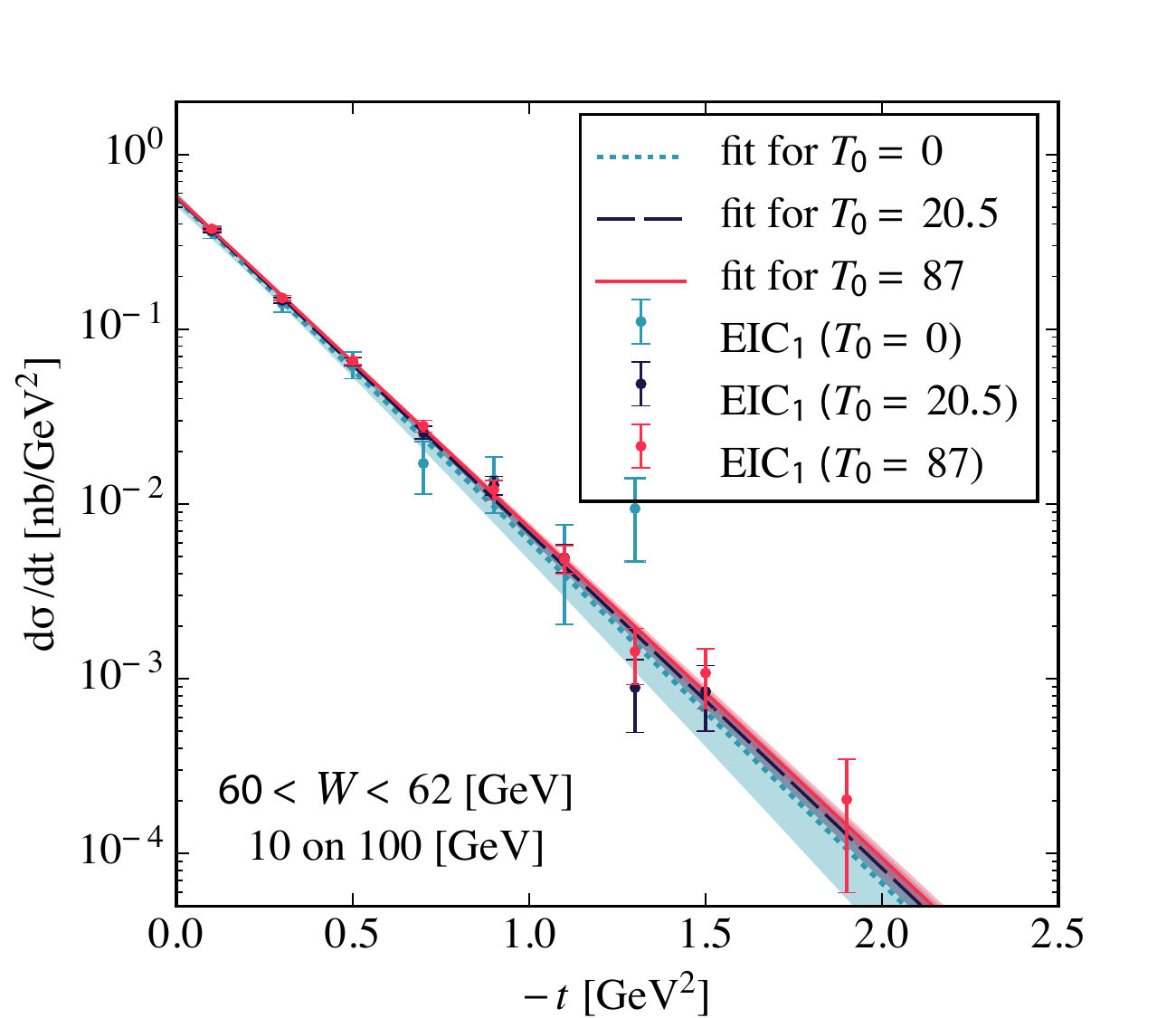}
\includegraphics[width=0.49\textwidth]{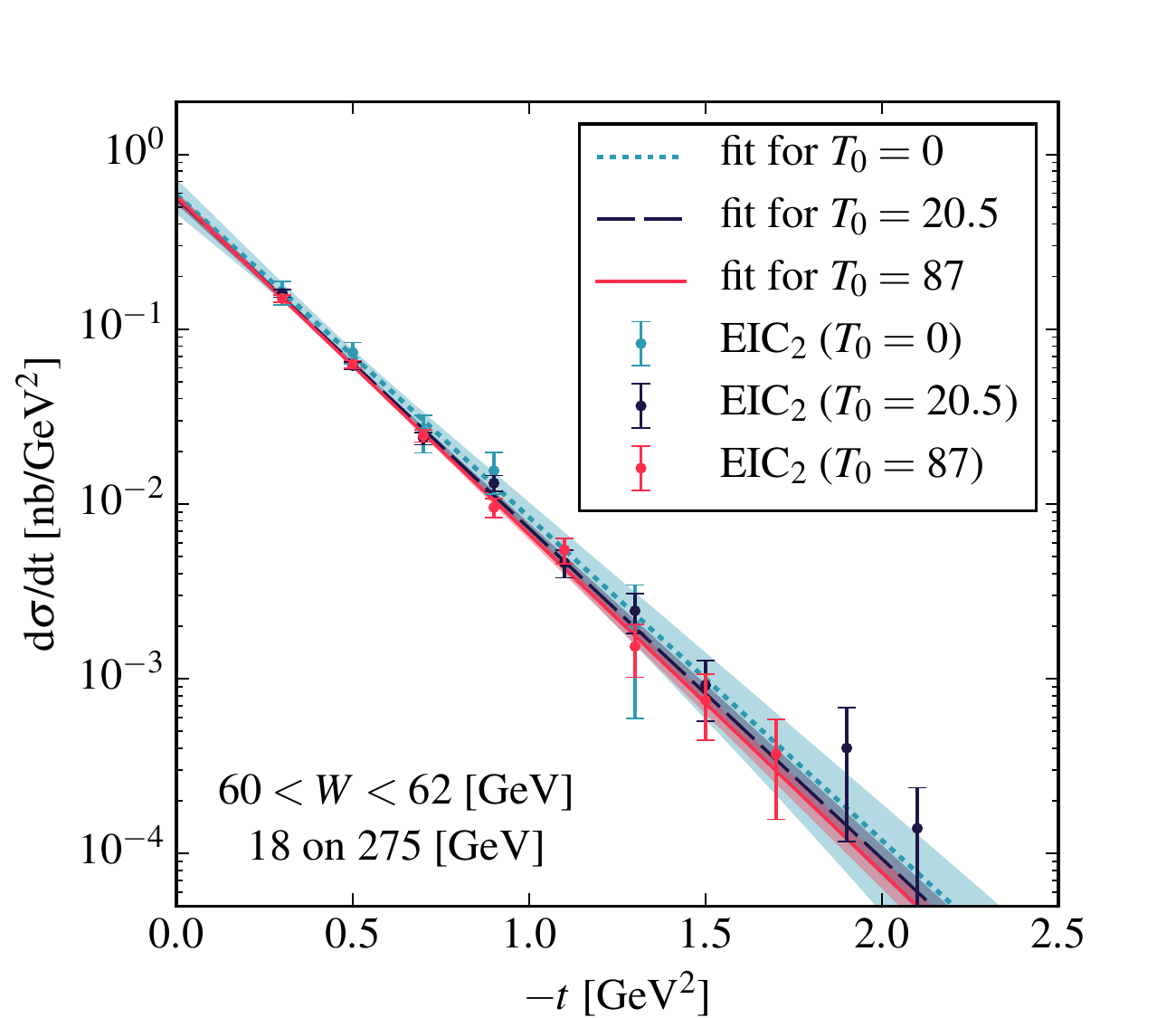}
\caption{$t$-dependence of the $\gamma p \to \Upsilon p$ differential cross section 
for different values of the subtraction constant $T_0 \equiv T_{\Upsilon p} (0)$ as indicated on the figure.
The EIC data points are simulated based on the theoretical elastic $\Upsilon$ photo-production cross section, assuming an exponential $t$-dependence. The bands represent the uncertainty propagated based on the simulated data points, 
assuming the two-parameter exponential fits.}
\label{fig:dsigmadt}
\end{figure*}

\begin{figure*}
\includegraphics[width=0.49\textwidth]{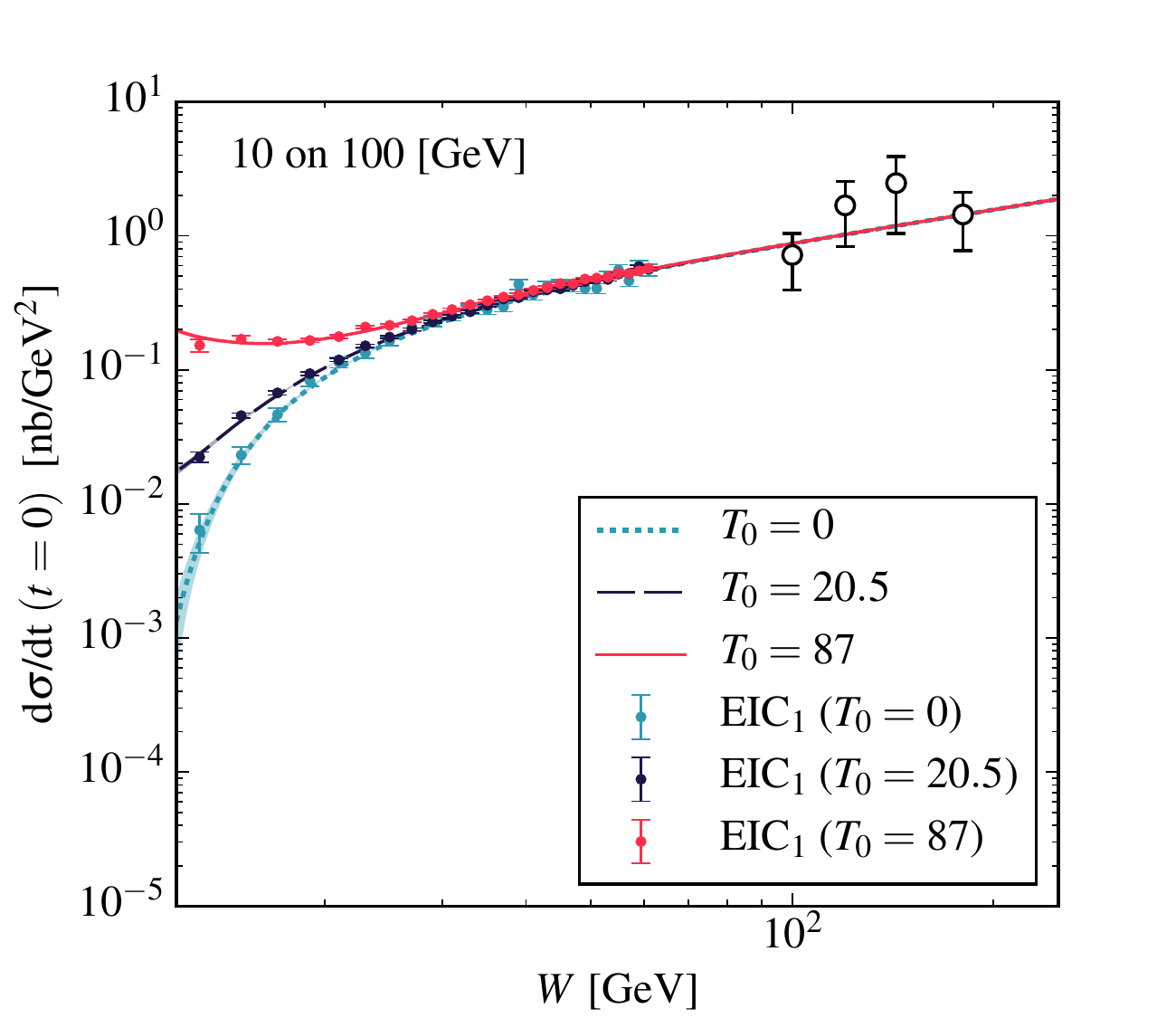}
\includegraphics[width=0.49\textwidth]{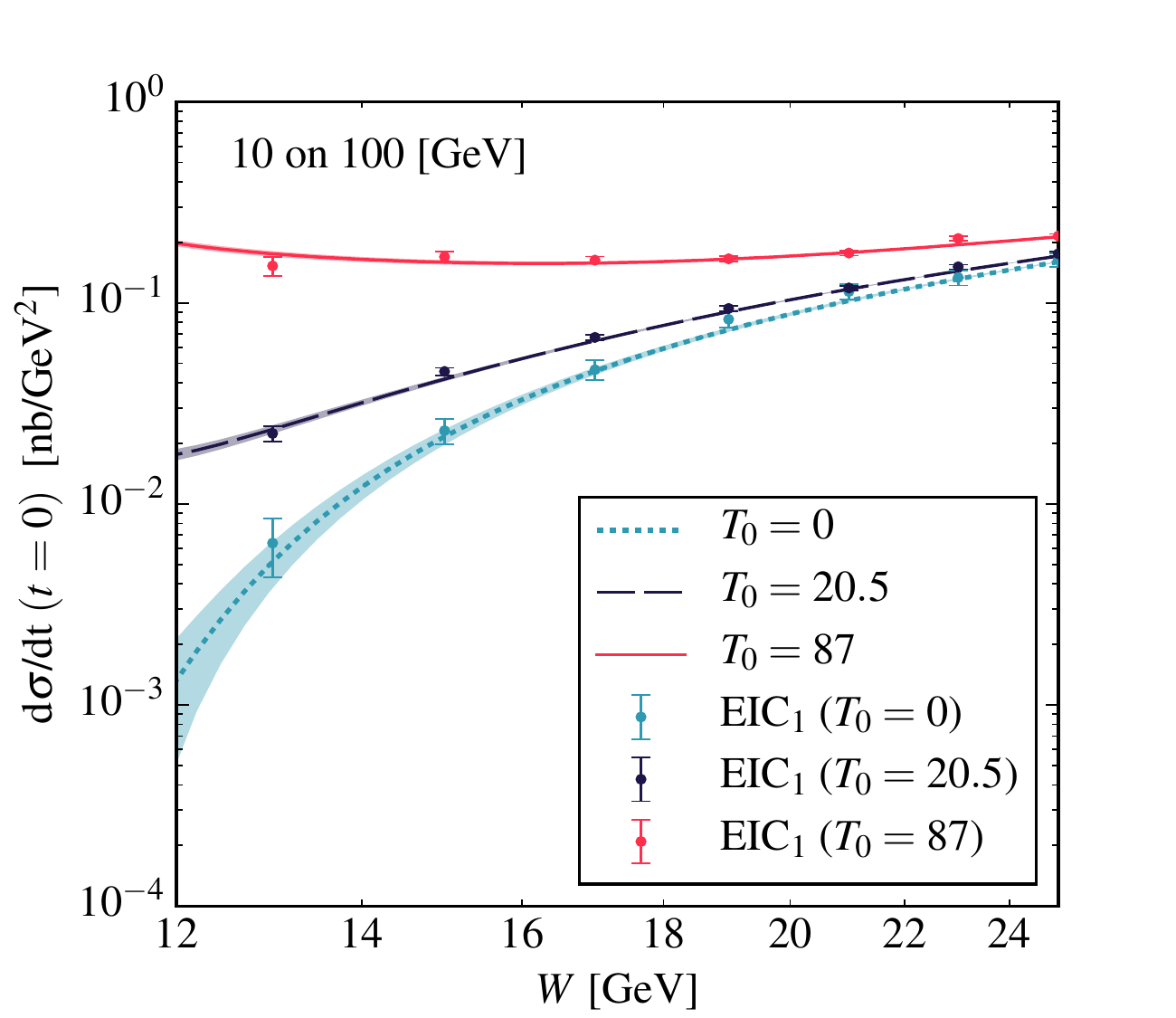} \\
\includegraphics[width=0.49\textwidth]{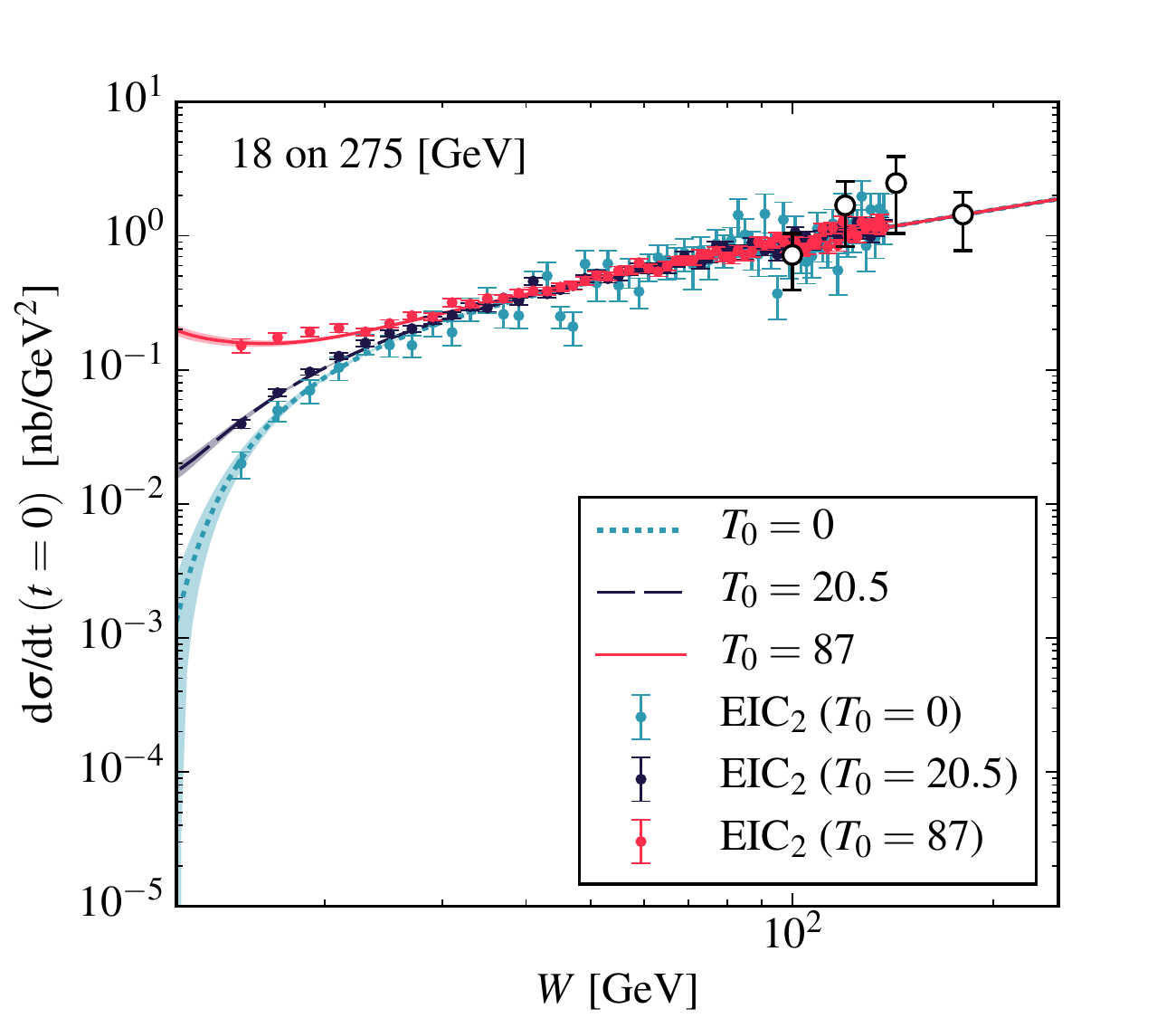}
\includegraphics[width=0.49\textwidth]{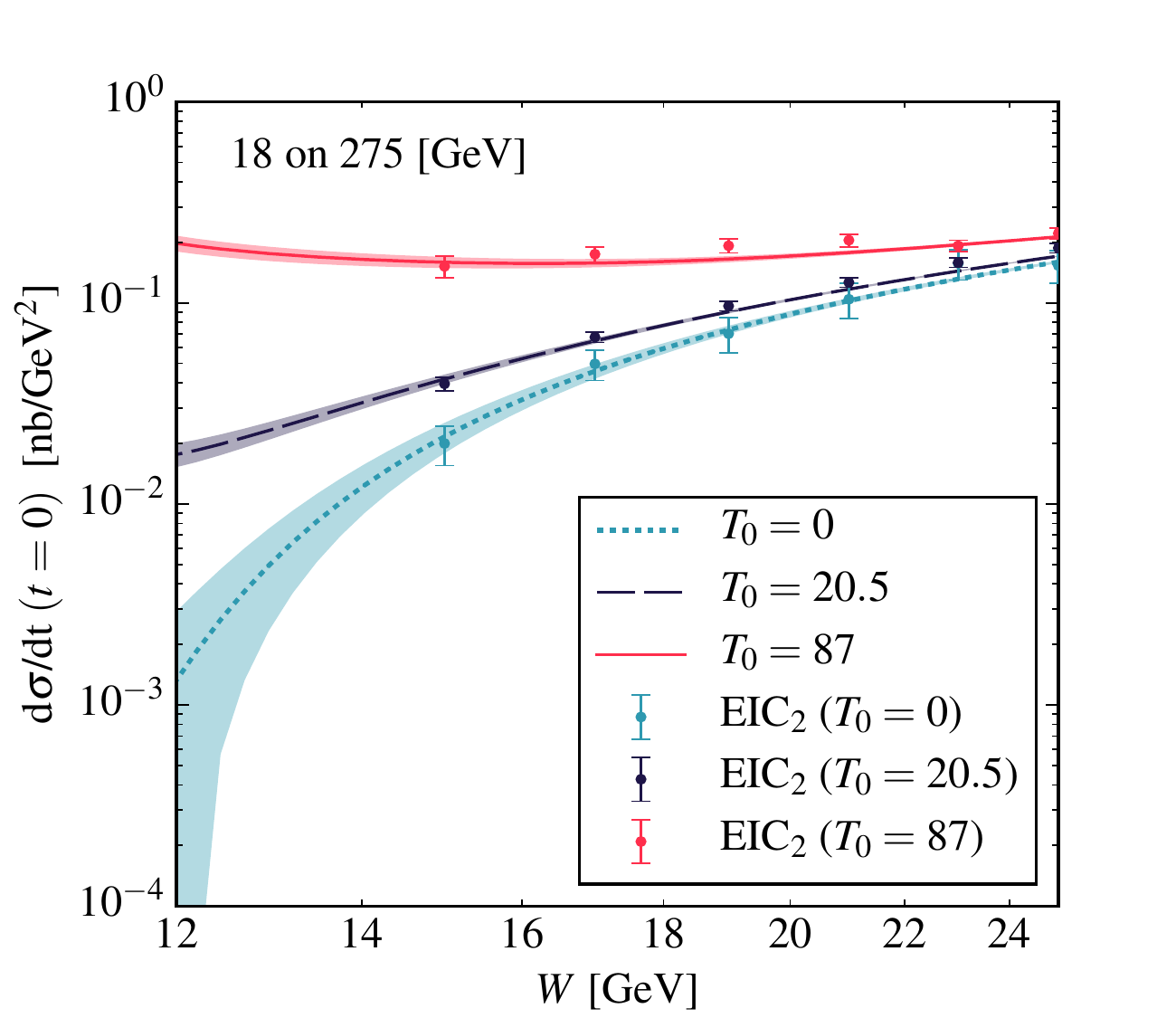}
\caption{$W$-dependence of the $\gamma p \to \Upsilon p$ differential cross section, extrapolated to $t=0$, 
for different values of the subtraction constant $T_0 \equiv T_{\Upsilon p} (0)$. The data points (open circles) are obtained from the elastic $\Upsilon$ photo-production cross 
section from HERA~\cite{Adloff:2000vm,Breitweg:1998ki,Chekanov:2009zz}
 by using the empirically measured slope parameter, using Eq.~(\ref{eq:brel}). The bands represent the uncertainty propagated based on the EIC simulated  data points, assuming a one-parameter fit of $T_0$. Upper (lower) panels are for EIC beam setting 1 (beam setting 2) respectively. Right panels give a more detailed view of the lower energy region $W < 25$~GeV.}
\label{fig:dsigmadt0}
\end{figure*}

We use the Argonne l/A-event generator~\cite{git:lager} to simulate a realistic event sample for the $\gamma p \to \Upsilon p$ process at the EIC, and refer to Appendix~\ref{apx-evgen} for more details on the exact implementation of the simulation. 
For our simulations, we assumed a total integrated luminosity of 100~fb$^{-1}$ for each of the settings, which corresponds to 116 days at 10$^{34}$cm$^{-2}$s$^{-1}$.
We simulated both the $\Upsilon \to e^+e^-$ and $\Upsilon \to \mu^+\mu^-$ decay channels, and only considered events where we fully detect the exclusive final state.
We assumed nominal EIC detector parameters in line with the EIC  white paper, where we have lepton detection for pseudo-rapidities between $-5 < \eta_l < 5$, and recoil proton detection for angles $\theta_p > 2$~mrad. 
Furthermore, we assumed we can reconstruct the scattered electron for 
$y = P.q/P.k$ between $0.01 < y < 0.8$, and we ensured a 
quasi-real regime by requiring that $Q^2 < 1$~GeV$^{2}$.

In Fig.~\ref{fig:dsigmadt}, we show the simulated results for the $t$-dependence of the $\gamma p \to \Upsilon p$ 
differential cross sections for different values of $W$,  
corresponding to the two EIC beam settings. 
In each case we consider the three scenarios for the subtraction constant $T_0$ discussed above. 
For the generated EIC data, the exponential $t$-slope $B$ is obtained through solution of Eq.~(\ref{eq:brel}). 
The error bands represent the uncertainty propagated based on the data points, assuming the two-parameter exponential fits of Eq.~(\ref{eq:bdef}). 
Only points generated up to $-t$ values of $2.5$ GeV are considered in the fits.

In Fig.~\ref{fig:dsigmadt0}, we show the $W$-dependence of the $\gamma p \to \Upsilon p$ differential cross sections extrapolated to t = 0 for the two EIC beam settings considered.  
We see that in the $W < 25$~GeV region, the precision that can be reached for beam setting 1 at the EIC will allow to clearly distinguish between the three scenarios for the subtraction constant, and allow to extract this parameter with a statistical precision of the order of a few percent, see Table~\ref{tab:scattlength}. 
With our nominal assumption of sufficient experimental resolution to reconstruct events down to $y \sim 0.01$, we can also distinguish between the three scenarios with beam setting 2, albeit with reduced statistical precision. 
This reach into the threshold region for high-energy beam settings is very sensitive to the experimental resolution to reconstruct scattered electrons at very low $y$:
a slight increase in the lower $y$ limit for beam setting
2 may make the measurement impossible, while a slight decrease in this limit improves the sensitivity to the same level as beam setting 1.
Furthermore, the EIC will allow to perform an independent fit of the 
$b_{\rm inel}$ parameter, which governs the low-energy behavior of the $\gamma p \to \Upsilon p$ forward differential cross section, according to Eq.~(\ref{eq:discx}). 
Beam setting 1 is particularly suited for this measurement, 
and beam setting 2 may also be sensitive to $b_{\rm inel}$ contingent on the precise experimental resolution of the EIC detector near the lower $y$ limit.
For the region $W > 25$~GeV, beam setting~2 will allow to connect the data with the existing HERA measurement and furthermore allow to perform an independent fit of the parameter $a_{\rm inel}$, which governs the high-energy behavior of the $\gamma p \to \Upsilon p$ forward differential cross section.

\section{Conclusion}

In this work, we have extended a previous dispersive study of $J/\psi$ photo-production to the case of $\Upsilon$ photo-production on a proton target, with the aim to extract the $\Upsilon$-p scattering length from future $\gamma p \to \Upsilon p$ experiments. The imaginary part of the $\Upsilon$-p forward scattering amplitude is constrained at high energies from existing HERA and LHC data for the $\gamma p \to \Upsilon p$ total and differential cross sections. 
Its real part is calculated through a once-subtracted dispersion relation, and the subtraction constant is proportional to the $\Upsilon$-p scattering length. 
As no data are available so far in the threshold region, we have  considered three scenarios for the subtraction constant: one of them corresponds to a zero value, the other to a value for the $\Upsilon$ study similar to that of the $J/\psi$ case, and a third where we estimate the subtraction constant by considering the $\Upsilon$ as a Coulombic bound state which interacts with the proton through its chromo-electric polarizability. 
Using these three scenarios, we have performed a feasibility study for $\Upsilon$ quasi-real photon ($Q^2 < 1~\rm{GeV}^2$) production experiments at an Electron-Ion Collider, and considered a low-energy and a high-energy beam setting. 
For our simulations, we have assumed a total integrated luminosity of 100~fb$^{-1}$ for each of the settings, assuming nominal EIC detector parameters in line with the EIC white paper.   
In both beam settings, the simulated data for the $\gamma p \to \Upsilon p$ cross section were found to clearly distinguish between the three considered scenarios for the subtraction constant.
The low-energy beam setting, accessing the range $12~{\rm{GeV}}\lesssim W \lesssim 60~{\rm{GeV}}$, was found to yield the higher statistical precision on the cross section. Furthermore, the high-energy beam setting, accessing the range $15~{\rm{GeV}}\lesssim W \lesssim 140~{\rm{GeV}}$, will allow to connect the EIC data with the existing HERA data, and thus provide an independent measurement of the high-energy behavior of the $\gamma p \to \Upsilon p$ forward differential cross section, further constraining the dispersive formalism. It is worth noting that our projection of the statistical error analysis shows that the $\Upsilon$-p scattering length can be extracted from such data with a statistical precision of the order of 2\% or less. The total experimental uncertainty of this determination will be dominated by the systematic error in the measurement of the absolute value of the differential cross section expected to be in the few percent range,  leading to a very potent determination of the scattering length.

Our work shows that an experimental program on $\Upsilon$ quasi-real photo-production at the EIC has the potential to provide a unique view on the gluonic van der Waals interaction in Quantum ChromoDynamics.

\section*{Acknowledgements}
The work of OG and MV was supported by the Deutsche Forschungsgemeinschaft (DFG, German Research Foundation),
in part through the Collaborative Research Center [The Low-Energy Frontier of the Standard
Model, Projektnummer 204404729 - SFB 1044], and in part through the Cluster of Excellence
[Precision Physics, Fundamental Interactions, and Structure of Matter] (PRISMA$^+$ EXC
2118/1) within the German Excellence Strategy (Project ID 39083149).
The work of SJ and ZEM is supported by the US DOE contract DE-AC02-06CH11357.

\appendix

\section{Event Generation\label{apx-evgen}}
In order to simulate a realistic event sample for $\Upsilon$ events at the EIC,
we added the formalism of this paper to the Argonne l/A-event Generator (\textsc{lager}) \cite{git:lager}.
\textsc{Lager} is a modular accept-reject generator capable of simulating both fixed-target and collider kinematics. Below we describe the model components used to obtain the event samples for this work.

\subsection{Differential electro-production cross section}

The differential cross section for the process ($e p \to e^\prime \gamma^* p \to e^\prime \Upsilon p$) can be written as,
\beq
\frac{d\sigma}{dQ^2dydt}(e p \to e^\prime \Upsilon p) = 
\Gamma_T(1+\epsilon R)
\frac{d\sigma}{dt}(\gamma^* p \to \Upsilon p),
\eeq
with transverse virtual photon flux $\Gamma_T$, virtual photon polarization $\epsilon$, and 
$R\equiv\sigma_L/\sigma_T$ parameterized as in Ref.~\cite{Martynov:2002ez},
\beq
R(Q^2) = \left(\frac{A M_\Upsilon^2 + Q^2}{A M_\Upsilon^2}\right)^{n_1} - 1.
\eeq
We use the values for parameters ($A$, $n_1$) as determined for $J/\psi$ production in Ref.~\cite{Fiore:2009xk}.
In order to estimate the unknown $Q^2$ dependence of the differential $\gamma^* p \to \Upsilon p$ cross section, we use the following factorized ansatz,
\beq
\frac{d\sigma}{dt}(\gamma^* p \to \Upsilon p) = D(Q^2) \frac{d\sigma}{dt}(\gamma p \to \Upsilon p),
\eeq
where for $D$ we assumed a dipole-like form-factor, similar to what is typically
done in a vector meson dominance model (VMD),
\beq
D(Q^2) = \left(\frac{M_\Upsilon^2}{M_\Upsilon^2 + Q^2}\right)^{n_2}.
\eeq
This formula deviates from its standard VMD form through the value for $n_2$, which was
tuned to optimally describe the $Q^2$ dependence for exclusive $\rho$ production in a wide range
of kinematic regions.
Note that this assumption has very little impact on the projections in this work, as we only consider quasi-real events.

\subsection{Differential cross section for $\Upsilon$ photo-production}

In order to determine the slope $B$ of the $t$-dependence of the differential cross section for 
the $\gamma p \to \Upsilon p$ process, we numerically solve the transcendental equation \eqref{eq:brel}.
Note that both  normalization $A$ and slope $B$ depend on the choice of the subtraction
constant $T_{\Upsilon p} (0)$, while the total integrated cross section 
$\sigma(\gamma p \to \Upsilon p)$ is independent of the subtraction constant.

\subsection{Angular dependence of the decay leptons}

We included both the $\Upsilon \to e^+e^-$ and $\Upsilon \to \mu^+\mu^-$ decay channels
in our simulation, using $s$-channel helicity conservation (SCHC) to describe the
angular distribution for a vector meson decaying into two fermions~\cite{Breitweg:1998nh,Chekanov:2002xi,Schilling:1973ag}:
\beq
\mathcal{W}(\cos\theta_\text{CM}) = 
\frac{3}{8}(1+r^{04}_{00}+(1-3r_{00}^{04})\cos^2\theta_\text{CM}),
\eeq
where we relate the spin-density matrix element $r^{04}_{00}$ to $R$ as,
\beq
R = \frac{1}{\epsilon}\frac{r^{04}_{00}}{1-r^{04}_{00}}.
\eeq

\end{document}

%% file: table.tex
$C_x$ & $(13.8\pm8.1)\times 10^{-3}$ & $18.7\pm2.3$ \\
$b_x$ & $1.27$ (fixed) & $3.53$ (fixed) \\
$a_x$ & $1.38\pm0.06$ & $1.2$ (fixed) \\

%% file: table2_eic1.tex
&$0$ & $\simeq 0$ & $\simeq 0$ \\
&$20.5\pm0.9$ & $0.016\pm0.001$ & $0.78\pm0.03$ \\
&$87\pm2$ & $0.066\pm0.001$ & $3.23\pm0.06$ \\

%% file: table2_eic2.tex
&$0$ & $\simeq 0$ & $\simeq 0$ \\
&$20.5\pm1.9$ & $0.016\pm0.001$ & $0.78\pm0.07$ \\
&$87\pm4$ & $0.066\pm0.003$ & $3.23\pm0.16$ \\